\documentclass[iop]{emulateapj}

\usepackage{hyperref}
\usepackage{subfigure}
\usepackage{amsmath}

\submitted{Accepted by The Astrophysical Journal}
\shorttitle{Dust Reverberation Lag of NGC 6418}
\shortauthors{Vazquez et al.}

\begin{document}

\title{Spitzer Space Telescope Measurements of Dust Reverberation Lags in the Seyfert 1 Galaxy NGC 6418}

\author{Billy Vazquez\altaffilmark{1},
        Pasquale Galianni\altaffilmark{2},
        Michael Richmond\altaffilmark{1},
        Andrew Robinson\altaffilmark{1},
        David J. Axon\textdagger\altaffilmark{1,3},
        Keith Horne\altaffilmark{2},
        Triana Almeyda\altaffilmark{1},        
        Michael Fausnaugh\altaffilmark{4},
        Bradley M. Peterson\altaffilmark{4,20},
        Mark Bottorff\altaffilmark{5},        
        Jack Gallimore\altaffilmark{9},
        Moshe Eltizur\altaffilmark{11},
        Hagai Netzer\altaffilmark{12},        
        Thaisa Storchi-Bergmann\altaffilmark{19},
        Alessandro Marconi\altaffilmark{7},
        Alessandro Capetti\altaffilmark{8},
        Dan Batcheldor\altaffilmark{6},        
        Catherine Buchanan\altaffilmark{10},
        Giovanna Stirpe\altaffilmark{18},
        Makoto Kishimoto\altaffilmark{13},
        Christopher Packham\altaffilmark{15},
        Enrique Perez\altaffilmark{16},
        Clive Tadhunter\altaffilmark{17},
        John Upton\altaffilmark{5} \&
        Vicente Estrada-Carpenter\altaffilmark{5}}
       
\affil{$^{1}$Physics Department, Rochester Institute of Technology, 84 Lomb Memorial Drive, Rochester, NY 14623-5603, USA}
\affil{$^{2}$SUPA, School of Physics and Astronomy, The University of St Andrews, North Haugh, St Andrews, KY169SS, UK}        
\affil{$^{3}$School of Mathematical and Physical Sciences, University of Sussex, Sussex House, Brighton, BN1 9RH, UK}
\affil{$^{4}$Department of Astronomy, The Ohio State University, 140 West 18th Avenue, Columbus, OH 43210-1173}
\affil{$^{5}$Department of Physics, Southwestern University, Georgetown, TX 78626}
\affil{$^{6}$Department of Physics and Space Sciences, Florida Institute of Technology, 150 W. University Blvd, Melbourne, FL 32901, USA}
\affil{$^{7}$Dipartimento di Fisica e Astronomia, Universita di Firenze, Via G.~Sansone 1, 50019, Sesto Fiorentino (Firenze), Italy}
\affil{$^{8}$INAF-Osservatorio Astronomico di Torino, Strada Osservatorio 20, 10025 Pino Torinese, Italy}
\affil{$^{9}$Department of Physics \& Astronomy, Bucknell University, 1 Dent Drive, Lewisburg, PA 17837}
\affil{$^{10}$University of Melbourne, 1—100 Grattan Street, Parkville, Victoria, 3010}
\affil{$^{11}$Department of Physics and Astronomy, University of Kentucky, Lexington, Kentucky 40506}
\affil{$^{12}$School Of Physics And Astronomy, Faculty of Exact Sciences, Tel Aviv University, Ramat Aviv, Israel}
\affil{$^{13}$Max Planck Institute for Astronomy, Königstuhl 17 69117 Heidelberg, Germany}
\affil{$^{15}$Department of Astronomy, University of Florida, Gainesville, FL 32611}
\affil{$^{16}$Instituto de Astrofisica de Andalucia, Glorieta de la Astronomía s/n, 18007, Granada, Spain}
\affil{$^{17}$Department of Physics and Astronomy, The University of Sheffield, Western Bank, Sheffield, South Yorkshire, S10 2TN, UK}
\affil{$^{18}$INAF - Osservatorio Astronomico di Bologna, Via Ranzani 1 40127, Bologna, Italy}
\affil{$^{19}$Departamento de Astronomia, Instituto de Física, Universidade Federal do Rio Grande do Sul, Campus do Vale, Av. Bento Goncalves 9500, 91501-970 Porto Alegre, RS, Brasil}
\affil{$^{20}$Center for Cosmology and AstroParticle Physics, The Ohio State University, 191 West Woodruff Avenue, Columbus, OH 43210}

\begin{abstract}
We present results from a fifteen-month campaign of high-cadence ($\sim$3 days) 
mid-infrared Spitzer and optical ($B$ and $V$) monitoring of the Seyfert 1 galaxy 
NGC 6418, with the objective of determining the characteristic size of the dusty torus in 
this active galactic nucleus (AGN). We find that the 3.6 $\mu$m and 4.5 $\mu$m flux 
variations lag behind those of the optical continuum by $37.2^{+2.4}_{-2.2}$ days 
and $47.1^{+3.1}_{-3.1}$ days, respectively. We report a cross-correlation time lag 
between the 4.5 $\mu$m and 3.6 $\mu$m flux of $13.9^{+0.5}_{-0.1}$ days. The lags indicate that the dust 
emitting at 3.6 $\mu$m and 4.5 $\mu$m is located at a distance $\approx 1$ light-month 
($\approx 0.03$\,pc) from the source of the AGN UV--optical continuum. 
The reverberation radii are consistent with the inferred lower limit to the
sublimation radius for pure graphite grains at 1800 K, but smaller by a factor of $\sim 2$
than the corresponding lower limit for silicate grains; this is similar to what has been found for 
near-infrared (K-band) lags in other AGN. The 3.6 and 4.5 $\mu$m reverberation radii fall above 
the K-band $\tau \propto L^{0.5}$ size-luminosity relationship by factors $\lesssim 2.7$ and 
$\lesssim 3.4$, respectively, while the 4.5 $\mu$m reverberation radius is only 27\% larger than 
the 3.6 $\mu$m radius. This is broadly consistent with clumpy torus models, in which 
individual optically thick clouds emit strongly over a broad wavelength range.
\end{abstract}
\keywords{galaxies: active --- galaxies: individual (NGC 6418) --- galaxies: nuclei --- galaxies: Seyfert}

\section{Introduction}
\label{sec:Introduction}

\begin{deluxetable*}{lccclll}
\tabletypesize{\scriptsize}
\tablecaption{Observations}
\tablewidth{0pt}
\tablehead{
\colhead{Telescope} & \colhead{start date} & \colhead{end date} & 
\colhead{\# obs} & \colhead{instrument} & \colhead{filter} & \colhead{aperture} 
}
\startdata
Liverpool Telescope & 08-06-2011 & 10-21-2012 & 64 & RATCam & Bessel B & 1.5" \\
Faulkes Telescope North & 08-10-2011 & 09-30-2012 & 60 & fs02 & Bessel B & 1.2" \\
SU Fountainwood 0.4-m& 05-19-2012 & 12-18-2012 & 48 & SBIG ST-8300 & Johnson-Cousins B/V & 3.5" \\
Spitzer & 08-01-2011 & 01-04-2013 & 170 & IRAC & ch1/ch2 & 1.8" \\
\enddata 
\label{obs_table}
\end{deluxetable*}

In the AGN unification paradigm, direct 
observation of the nucleus is blocked by a toroidal structure of dusty 
molecular gas for a range of viewing angles \citep[e.g.,][]{Antonucci:1993}. 
As this dust absorbs UV--optical radiation from the accretion disk and 
re-emits in the infrared (IR), this structure is also thought to be the dominant 
source of IR radiation in most AGN. Understanding this obscuration of the central 
engine is therefore important to understanding the physical processes operating in 
AGN and more generally, their role in galaxy evolution. 

The observational evidence \citep{Antonucci:1993, Jaffe:2004, Tristram:2007}, indicates that 
the obscuring structure is geometrically and optically thick, although a warped thin disk 
that extends throughout the host galaxy has also been proposed\citep{Sanders:1989}.
The conventional picture is that of a compact, but geometrically thick, torus of 
optically thick molecular clouds with a size of a few parsecs
\citep{Antonucci:1985,Krolik:1988,Pier:1992}. Models in which the vertical thickness 
is supported by large random velocities due to elastic collisions 
between clouds\citep{Krolik:1988}, or by IR radiation 
pressure \citep{Pier:1992, Krolik:2007}, or by turbulence induced by 
supernovae \citep{Wada:2002, Schartmann:2009} have been explored.
In an alternative class of models, the dusty material is not part of an essentially 
static torus, but is rather embedded in
an outflowing hydromagnetic wind launched from the accretion 
disk \citep[e.g.,][]{Blandford_Payne:1982,Emmering:1992,Bottorff:1997,Elitzur:2006,Dorodnitsyn:2012}. 

Dust radiative transfer models for the torus broadly reproduce the IR spectral 
energy distribution (SED) of AGN. Of necessity, early radiative transfer models assumed
smooth density distributions \citep[e.g.][]{Pier:1993,Granato:1994, Efstathiou:1995}, but
more recently, models for clumpy dust distributions have been developed
\citep[e.g.,][]{Nenkova:2002,Dullemond:2005,Honig:2006,Schartmann:2008,Nenkova:2008a,Nenkova:2008b}. 
These ``clumpy torus'' models are more successful in reproducing certain details of the 
SED such as, for example, the strength of the 10 $\mu$m silicate feature \citep{Nikutta:2009, Nenkova:2008b}. 

The torus is too small to be directly imaged by any existing single telescope. 
Some constraints on its size and structure can be inferred from SED-fitting using radiative transfer models
\citep[e.g.,][]{Nenkova:2008b, Mor:2009, Honig:2010, RamosAlmeida:2011, AlonsoHerrero:2011}, but  
there are many theoretical and observational uncertainties which obfuscate the results. 
Other methods are therefore required, the two most important being  
reverberation mapping and, for relatively close objects,  IR interferometry. \\

Following the seminal work of \citep{Blandford:1982}, the reverberation mapping technique has
been well developed and extensively applied to studies of the broad emission line region (BLR). 
Time series analysis of the response of the broad emission lines to variations in the UV or optical continuum
(as proxies for the AGN ionizing continuum)
has revealed the characteristic size of the BLR in about 50 AGN, enabling estimates of black hole 
masses and Eddington ratios 
\citep[][and references therein]{Peterson:1993,Peterson:2006,Gaskell:2009,Galianni:2013,Du:2014}. 
It has also been determined that the BLR follows a size--luminosity relationship of the 
form $R\propto L^{1/2}$ \citep{Peterson:2004, Greene:2010, Bentz:2013},  

Near-IR (K-band) versus optical (V-band) reverberation lags 
have been measured for around 20 Seyfert 
galaxies \citep{Oknyanskij:2001,Minezaki:2004,Suganuma:2006,Koshida:2009,Koshida:2014}.
As dust grains emitting in the K-band have temperatures close to the sublimation temperature ($\sim 1200-1800$\,K,
depending on grain composition), these lags are thought to represent
 the inner radius of the torus. The K-band reverberation lags are found to be larger than those of the BLR,
while following a similar $R\propto L^{1/2}$ size-luminosity relation, 
implying that the BLR is bounded by the dust distribution, consistent with the central idea 
of the AGN unification scheme. \\

The inner regions of several bright, nearby Seyfert galaxies have been directly studied using 
near-IR (K-band) interferometry \citep{Swain:2003,Kishimoto:2009b,Pott:2010,Kishimoto:2011,
Weigelt:2012}. The effective ring radii derived from the observed visibilities scale approximately as $L^{1/2}$,
and are comparable with or slightly larger than the radii derived from reverberation lags \citep{Kishimoto:2011}. 
Since \citet{Jaffe:2004}'s pioneering study of the archetypal Seyfert 2 galaxy, NGC\,1068, mid-IR ($8-12\mu$m)
interferometric observations have also been obtained for $\approx 20$ 
AGN \citep[e.g.,][]{Tristram:2007, Burtscher:2009, Kishimoto:2009a, Tristram:2009, Honig:2013} In 
a recent analysis of the available data, \citet{Burtscher:2013} find that while the mid-IR source 
size scales with luminosity in a manner similar to that seen in the near-IR, the inferred size is more 
than an order of magnitude larger than the measured K-band size and the scatter is quite large.

Here we report initial results from a mid-IR (3.6 $\mu$m and 4.5 $\mu$m) reverberation-mapping 
campaign using the Spitzer Space Telescope in its ``warm mission". 
Our motivation is to probe the dust distribution at spatial scales intermediate between 
the innermost regions probed by the 
K-band observations and the outer, cooler regions probed by  mid-IR interferometry.
Furthermore, variability at 3.6 $\mu$m and 4.5 $\mu$m should be less susceptible than 
the 2.2 $\mu$m $K$-band to complicating effects such as dust 
sublimation \citep{Minezaki:2004, Kishimoto:2013}, or contamination by variable accretion 
disk  emission \citep{Tomita:2006, Kishimoto:2007}. 
During a 2-year campaign, we monitored a sample of 12 Seyfert 1 AGNs at cadences of 3 and 30 days 
during the first and second year, respectively. 
We selected our targets based on their proximity ($z < 0.4$) and 
their location near one of Spitzer's continuous viewing zones.
We obtained $B$ and $V$ images of 
the targets over the same period using
 the Liverpool Telescope, the Faulkes Telescope North and the 
Southwestern University 0.4-m telescope.

In this work we describe our analysis
of the first 17 months of measurements of the Seyfert 1 NGC 6418 
\citep{Veron-Cetty:2006},
a Hubble type Sab galaxy \citep{Nair:2010} 
with an apparent magnitude $g = 14.87$
at a redshift of $z=0.0285$ \citep{Ahn:2014}.
It is classified spectroscopically as a Seyfert 1 on the basis of a strong, broad H$\alpha$ emission line, 
but it is otherwise dominated by the stellar continuum (see \citealt{Remillard:1993}, who described 
it as an ``embedded'' AGN). Nevertheless, it is also an X-ray
source with a 0.1-2.4 keV luminosity of $L_X= 10^{42.26}$ erg/s \citep{Anderson:2007}. We selected NGC 6418 
out of our sample due to its larger than average variations in the Spitzer channels for the first year of data; 
the result of the analysis of the other targets will be presented in a future publication.

We present our observations 
and describe our methods for measuring the 
light curves in Section~\ref{sec:Observations}.
In Section~\ref{sec:TimeSeriesAnalysis} we describe the 
time series analysis technique 
that was used to extract the time lags between the 3.6 $\mu$m, 
4.5 $\mu$m and optical light curves. 
We discuss the implications of our results 
in Section~\ref{sec:Discussion},
and present our conclusions in 
Section~\ref{sec:Summary}.
Details of our photometric measurements and a comparison of
two methods for determining time lags can be found in
the appendices.

\section{Observations}
\label{sec:Observations}

We will discuss the mid-infrared and optical observations 
separately.
See appendix \ref{app:a} for a detailed discussion of 
our photometric analysis. 

\subsection{Mid-Infrared}
We monitored 12 AGN using the Infrared Array Camera (IRAC) aboard the Spitzer Space Telescope 
for a period of approximately 
2 years during Cycles 8 (program 80120) and 9 (program 90209) 
of the ``warm'' mission. All objects were 
observed in both IRAC Channel 1 (3.6 $\mu$m) and Channel 2 (4.5 $\mu$m). 
During Cycle 8, repeated observations of each object
were obtained at intervals of 3 days. 
In Cycle 9, a longer cadence was used, with 30 day intervals between observations.
Here we report results from the Cycle 8 high-cadence monitoring of NGC 6418.
Images of this object were obtained every 3 days from 2011 Aug to 2013 Jan, 
except for a 30-day gap in 2011 Dec. Each image had an exposure of 10 seconds.  
All the resulting IRAC images were mosaiced using MOPEX 
\citep{MOPEX:2008} directly from the Basic Calibrated Data (BCD) level 1 products. 
Photometry was extracted from the BCD mosaics generated by the MOPEX standard pipeline,
as described in Section \ref{sec:Photometry}.

\subsection{Optical}
Contemporaneous optical monitoring was performed in the $B$ and $V$ bands 
with three ground-based telescopes:
Bessel $B$ images were obtained with the 2-m Liverpool Telescope (LT) on La Palma and the 
2-m Faulkes Telescope North (FTN) on Maui; Johnson-Cousins 
$B$ and $V$ images were obtained with the 0.4-m telescope at Southwestern University's (SU) 
Fountainwood Observatory (see Table \ref{obs_table}). 
It was not possible to coordinate these observations with each other 
or with the Spitzer observations, but 
together they approximately span the time period 
covered by the Spitzer campaign except during November 2011, 
when NGC 6418 was unobservable from the ground.
The start and end dates
of the observations with each telescope are given in Table~\ref{obs_table}.

The exposure times for the optical observations range from 60 to 
180 sec. 
Dark/bias subtraction 
and flat-field division of all images from SU
were performed using the 
XVISTA software package \citep{Treffers:1989}. 
Images from the RATCam instrument at
the LT were biased subtracted and flat fielded by an 
automatic pipeline \citep{Liverpool:2004},
as were images taken by the FTN.
When more than a single exposure per night was available from LT and 
FTN, we stacked and registered the images using MATCH, an implementation of 
the star matching algorithm of \cite{Tabur:2007}, and the XVISTA package. We then 
extracted photometry from the stacked image.

Hereafter, we refer to the light curve compiled from the LT and FTN observatories 
as the combined optical light curve. The SU observations are used to determine the AGN/Host ratio. 
The mean flux densities measured within the aperture in table \ref{obs_table} for all bands 
are tabulated in table\ref{mean_fluxes}. These flux densities are not host subtracted.

\begin{deluxetable}{lccc}
\tabletypesize{\scriptsize}
\tablecaption{Mean Flux Density}
\tablewidth{0pt}
\tablehead{
\colhead{name} & 
\colhead{mean flux density}
}
\startdata
3.6 $\mu$m & 3.62 mJy \\[4pt]
4.5 $\mu$m & 3.54 mJy \\[4pt]
SU B band & 0.53 mJy \\[4pt]
SU V band & 1.48 mJy \\[4pt]
LT B band & 0.50 mJy \\[4pt]
FTN B band & 0.54 mJy \\[4pt]
\enddata 

\label{mean_fluxes}
\end{deluxetable}

\subsection{Photometry}
\label{sec:Photometry}

\begin{figure}
\centering
\includegraphics*[scale=.75]{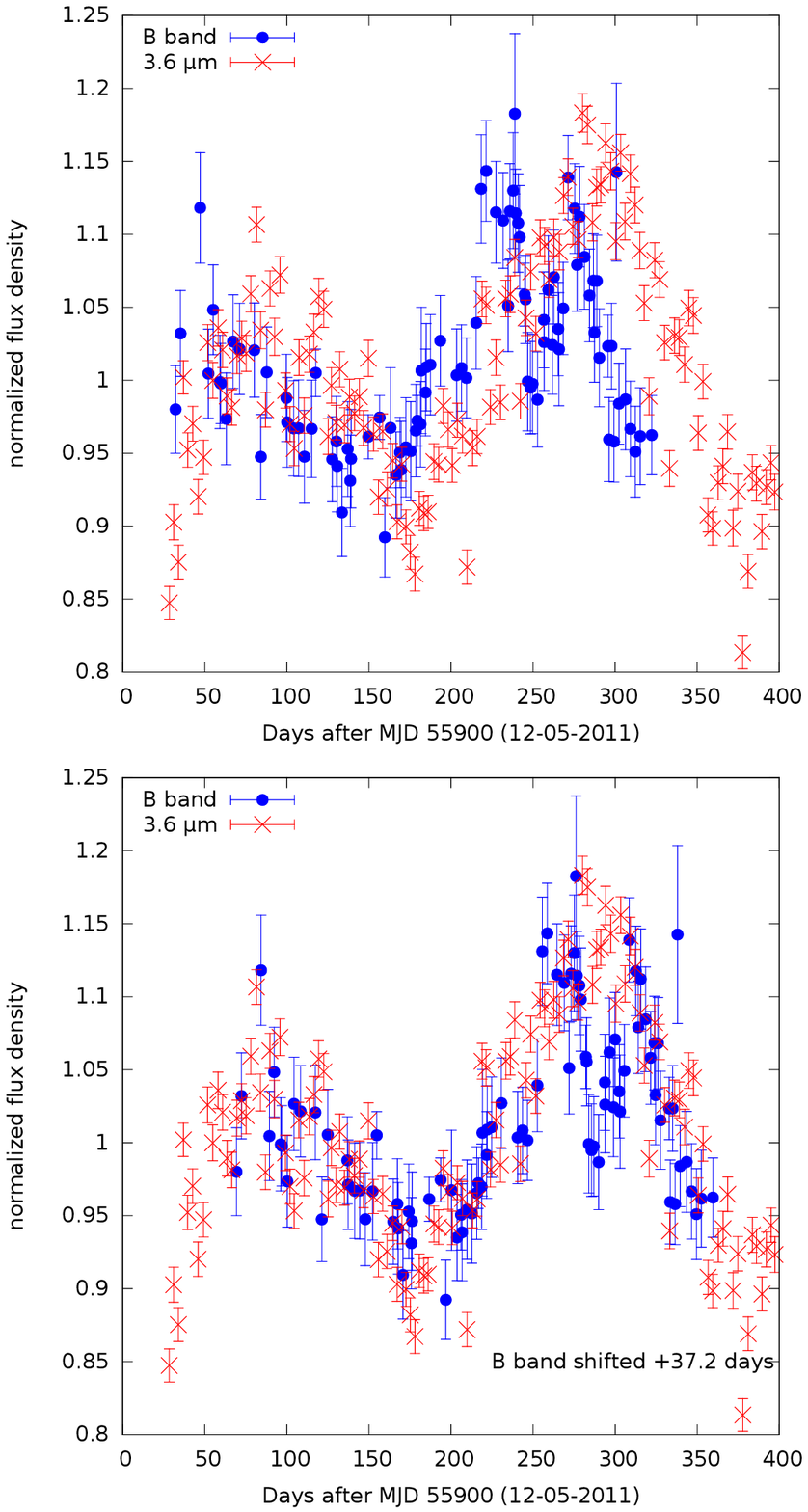}
\caption{Spitzer 3.6 $\mu$m and the combined $B$ band optical data light curves. The error bars of the 
3.6 $\mu$m and the combined $B$ band optical light curves are the uncertainties reported 
by MOPEX and the image differencing solution, respectively. The bottom panel shows 
the combined optical light curve shifted by +37.2 days.}
\label{fig:opt_ch1}
\end{figure}

\begin{figure}
\centering
\includegraphics*[scale=.75]{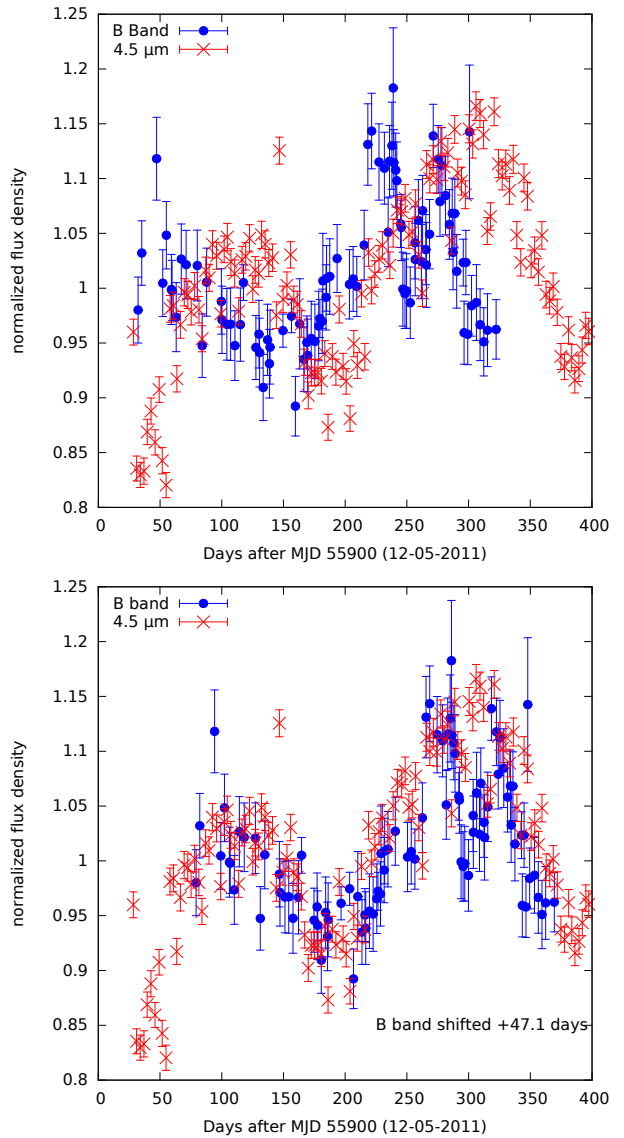}
\caption{Spitzer 4.5 $\mu$m and the combined $B$ band optical data light curves. The error bars of the 
4.5 $\mu$m and the combined $B$ band optical light curves are the uncertainties reported by MOPEX 
and the image differencing solution, respectively. The bottom panel shows the 
combined optical light curve shifted by +47.1 days.}
\label{fig:opt_ch2}
\end{figure}

The photometric analysis proceeds in two stages for the SU dataset : 
in the first we measure instrumental magnitudes for each object 
(the target plus comparison stars) in all exposures; 
in the second the measurements from all exposures in a 
given passband are combined and the 
measured instrumental magnitudes are subjected to 
inhomogeneous ensemble photometry \citep{Honeycutt:1992}. 
For a detailed discussion of these steps see
appendix \ref{app:a}. 
The LT and FTN datasets are reduced using image 
differencing \citep{Alard:2000} and references therein. 
The combined optical and Spitzer light curves are shown
in flux density, normalized to the mean, in figures \ref{fig:opt_ch1} and \ref{fig:opt_ch2}. 
The light curves are also shown after applying a shift equal to the time lag computed 
by the cross-correlation analysis (Sec. \ref{sec:TimeSeriesAnalysis}). 
In figures \ref{fig:opt_ch1} and \ref{fig:opt_ch2} the time lag shifts are $37.2$ and 
$47.1$ days for the Spitzer's $3.6 \mu$m/optical and Spitzer's $4.5\mu$m/optical, 
respectively.

The optical and infrared curves all show 
clear variations with similar features on timescales of 
$\sim 100$ days, 
but with the variations in the infrared lagging behind those in the optical.

\section{Time Series Analysis}
\label{sec:TimeSeriesAnalysis}

The reverberation lag, $\tau$, between the driving optical continuum 
variations and those of the responding IR 
emission gives the characteristic size of the IR emitting region.
The lag can be determined by cross-correlating the two light curves. 
The application of this technique to the broad emission line
variability of AGN (``reverberation mapping'') is well developed \citep{Gaskell:1986,
Gaskell:1987,Edelson:1988,Maoz:1989,Koratkar:1991} 
and has been widely used to measure the size of the broad line region   
(e.g., \citealt{Peterson:2004}; see \citealt{Peterson:2001} for a tutorial). 
As already noted, it has 
also been applied to optical and $K$-band
light curves in order to determine the inner radius of the torus \citep{Suganuma:2006, 
Oknyanskij:2006, Koshida:2009}. \\
 
We performed cross-correlation analyses for three pairs of data sets: 
3.6 $\mu$m versus combined optical,  
4.5 $\mu$m versus combined optical, and 4.5 $\mu$m versus 3.6 $\mu$m. 
The time series analysis was performed between the dates of MJD 55900 (12-05-2011)
and MJD 56300 (1-08-2013). This time span was selected due to the significant optical and 
IR variations of the light curves and because there were no large gaps in coverage.
For a comprehensive
and detailed analysis of individual datasets see appendix \ref{app:b}.
For each pair, the cross-correlation function (CCF) was computed
using a lag step size of 1 day. 
The optical observations were not synchronized with the Spitzer observations and 
are typically separated by irregular intervals. 
On the other hand, the Spitzer light curves are for the most part 
more evenly and densely sampled than the optical measurements. Therefore, in order 
to compute the IR--optical CCFs, we generate IR data points 
corresponding to the optical observations by interpolating within the 
Spitzer light curves. For examples of the CCFs computed for the 3 pairs of light curves
see appendix \ref{app:c}.  \\

The maximum of the CCF yields the lag, $\tau$, between the two light curves. 
However, the maximum is not always well defined, since 
computed CCFs typically exhibit a broad peak 
(see appendix \ref{app:c}) and structure in the wings (at large positive or
negative lags), which can influence the calculation of the centroid or mean. 
A common approach is to calculate the centroid of the CCF using a subset 
of points whose correlation coefficients exceed a certain value;
for example, 80\% of the peak value \citep{Peterson:2001}. 
Here, we use a different method in which we fit a cubic spline to the 
CCF and use it to set a threshold for the minimum correlation coefficient. This minimum 
correlation coefficient is defined as:

\begin{equation}
CC_{min}=\max(CCF(\tau))-2\sigma(CCF_{fit}(\tau)-CCF(\tau))
\end{equation}

where $CC_{min}$ is the minimum correlation coefficient, $CCF(\tau)$ is the 
cross-correlation function and $\sigma(CCF_{fit}(\tau)-CCF(\tau))$ is the 
standard deviation of the difference between the fitted and actual CCF value. 
The CCF centroid is computed using only values exceeding $CC_{min}$. For more details
see appendix \ref{app:c}.  \\

\begin{figure*}
\plotone{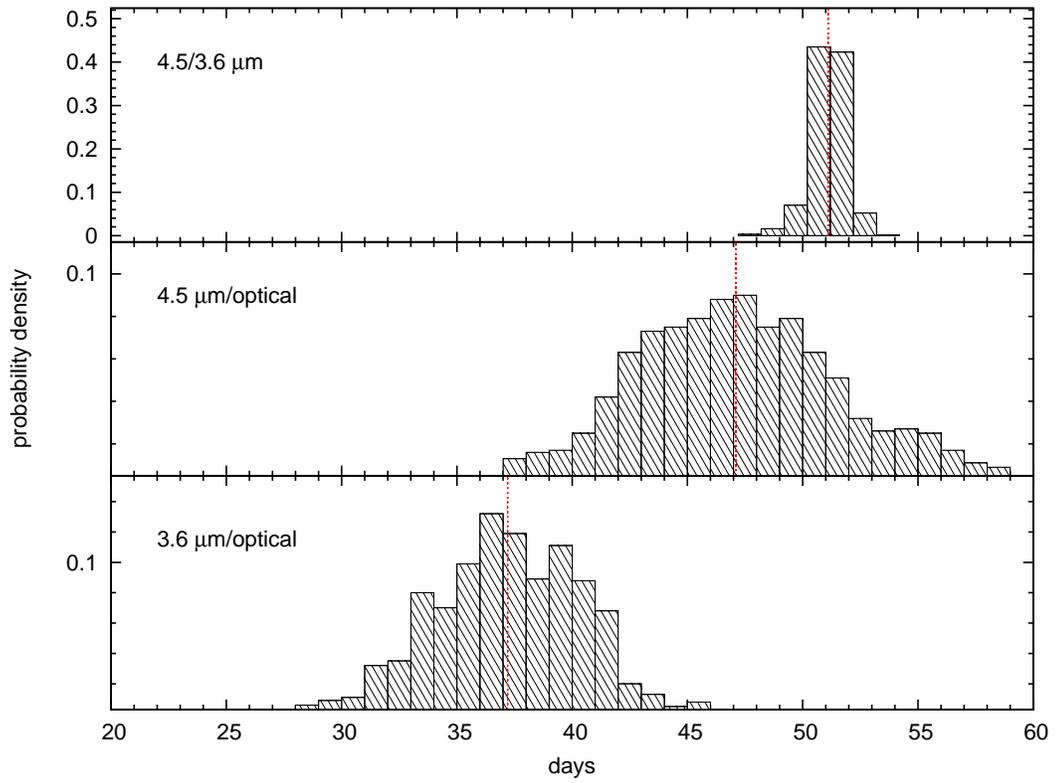}
\caption{Cross-correlation centroid distributions (CCCDs) for 
         3.6 $\mu$m versus 4.5 $\mu$m (top), 
         4.5 $\mu$m versus optical (middle), 
         3.6 $\mu$m versus optical (bottom).
         We have shifted the 3.6 $\mu$m versus 4.5 $\mu$m
         CCCD by 37.2 days, approximately the time lag between of the 
         3.6 $\mu$m and optical light curves, since, in principle, 
         we expect its peak to coincide with that of the 
         4.5 $\mu$m versus optical CCCD.}
\label{fig:CCCD_FR}
\end{figure*}

To estimate the uncertainty on the CCF lags, we used the 
cross-correlation centroid distribution (CCCD) method \citep{Gaskell:1987,Maoz:1989,Peterson:1998}, 
generating 1000 random realizations of the light curves. The CCCDs for the 3 pairs of data 
sets are shown in Figure \ref{fig:CCCD_FR} and
the derived lags are listed in Table \ref{codes} and in appendix \ref{app:b}. The lag is taken to be 
the median of the distribution and the uncertainty is given by the interquartile range.  
The CCCDs for the 3.6 $\mu$m versus optical, 
4.5 $\mu$m versus optical and 3.6 $\mu$m versus 4.5 $\mu$m light curves yield lags
of $37.2^{+2.4}_{-2.2}$ days ($31.2^{+2.0}_{-1.9}\times10^{-3}\;\text{pc}$), 
$47.1^{+3.1}_{-3.1}$ days ($39.5^{+2.6}_{-2.6}\times10^{-3}\;\text{pc}$),
and $13.9^{+3.7}_{-3.8}$ days ($11.7^{+0.4}_{-0.1}\times10^{-3}\;\text{pc}$), 
respectively.  \\

\begin{deluxetable}{lccc}
\tabletypesize{\scriptsize}
\tablecaption{Comparison of cross-correlation methods}
\tablewidth{0pt}
\tablehead{
\colhead{name} \vspace{-0.2cm} & 
\colhead{3.6 $\mu\text{m}$-Optical} & 
\colhead{4.5 $\mu\text{m}$-Optical} & 
\colhead{3.6 $\mu\text{m}$-4.5 $\mu\text{m}$} \\ \\
\colhead{} & 
\colhead{($\text{lag(day)}\pm\;\delta$)} & 
\colhead{($\text{lag(day)}\pm\;\delta$)} & 
\colhead{($\text{lag(day)}\pm\;\delta$)}
}
\startdata
Peterson et al. & 36.7$\pm$3.4 & 48.6$\pm$3.7 & 14.6$\pm$6.0 \\[4pt]
Zu et al. & $\text{40.4}^{+0.7}_{-6.5}$ & $\text{49.5}^{+1.2}_{-4.7}$ & $\text{13.2}^{+5.8}_{-2.9}$ \\[4pt]
Vazquez et al. & $37.2^{+2.4}_{-2.2}$ & $47.1^{+3.1}_{-3.1}$ & $13.9^{+0.5}_{-0.1}$ \\
\enddata 

\label{codes}
\end{deluxetable}

For comparison, we also analyzed our data following the slightly different cross-correlation 
methods described by \citet{Peterson:2004} and \citet{Zu:2011}. The results are compared in 
Table \ref{codes}.
We find that all methods yield results which are consistent within the uncertainties 
for all three pairs of light curves.

\section{Discussion}
\label{sec:Discussion}

The dusty torus absorbs UV/optical radiation from the accretion disk and 
re-emits it as infrared radiation \citep{Telesco:1984, Sanders:1989}. Variability in the 
accretion disk emission results in corresponding variations in the dust IR emission,
but with a delay due to differing light travel times between the source, various 
points in the torus and the observer. The lags between the optical continuum light curve 
and the IR light curves can therefore be interpreted as measures of the distance 
from the source to the dust clouds that predominantly emit the 3.6 $\mu$m and 4.5 $\mu$m
radiation. Our results indicate these clouds are located at a distance $\approx 1$ 
light-month ($\approx 0.03$\,pc) from the source of the AGN UV--optical continuum. 
However, the two Spitzer bands have significantly different lags, with the 
4.5 $\mu$m--optical lag being longer by $9.9\pm3.9$ days. The lag between the 
4.5 $\mu$m and 3.6 $\mu$m light curves is $13.9\pm0.5$ days and is consistent with this 
difference. This implies that the clouds producing the bulk of the 4.5 $\mu$m emission
are about 10 light-days ( $\sim 27$\%) further from the UV--optical continuum source.\\ 

In most models, the innermost radius of the torus is taken to be the dust sublimation 
radius which, for a typical ISM dust composition with silicate grains of average size, is
\citep{Barvanis:1987,Nenkova:2008b}

\begin{equation}
R_{d,Si} \simeq 1.3\left(\frac{L_{bol}}{10^{46}\;\text{erg}s^{-1}}\right)^{1/2}\left(\frac{1500\;\text{K}}{T_{sub}}\right)^{2.6}\;\text{pc}\label{sub_rad}
\end{equation}

where $L_{bol}$ is the bolometric luminosity of the AGN and $T_{sub}$
is the dust sublimation temperature. 

However, many broad-line AGN exhibit a distinct near infrared ''bump'', 
peaking around $2-4 \mu$m, which has 
a black body temperature $T\gtrsim 1000$\,K 
\cite[e.g.][]{Edelson:1986, Barvanis:1987, Rodriguez:2006, Riffel:2009a, Riffel:2009b}.
This feature often dominates the NIR and it has been found that it cannot 
be reproduced by torus models alone in fits to the infrared spectral energy 
distribution (SED); instead, one must add a
separate hot ($T\sim 1400$\,K) black body component. 
The latter has been attributed to hot pure graphite dust located within 
the torus \citep{Mor:2009, Mor:2011}, and  
\cite{Mor:2012} have modeled this component as dust embedded in the outermost BLR, 
between the sublimation radius for pure-graphite grains,

\begin{equation}
R_{d,C} \simeq 0.5\left(\frac{L_{bol}}{10^{46}\;\text{erg }s^{-1}}\right)^{1/2}\left(\frac{1800K}{T_{sub}}\right)^{2.8}\;\text{pc}\label{sub_rad_graphite}
\end{equation}

and the torus inner radius as given by equation~\ref{sub_rad}. The hot dust spectrum 
computed by \citeauthor{Mor:2012} suggests that this hot graphite dust 
contributes significant luminosity at 3.6 $\mu$m and 4.5 $\mu$m. 

\begin{figure*}
\plotone{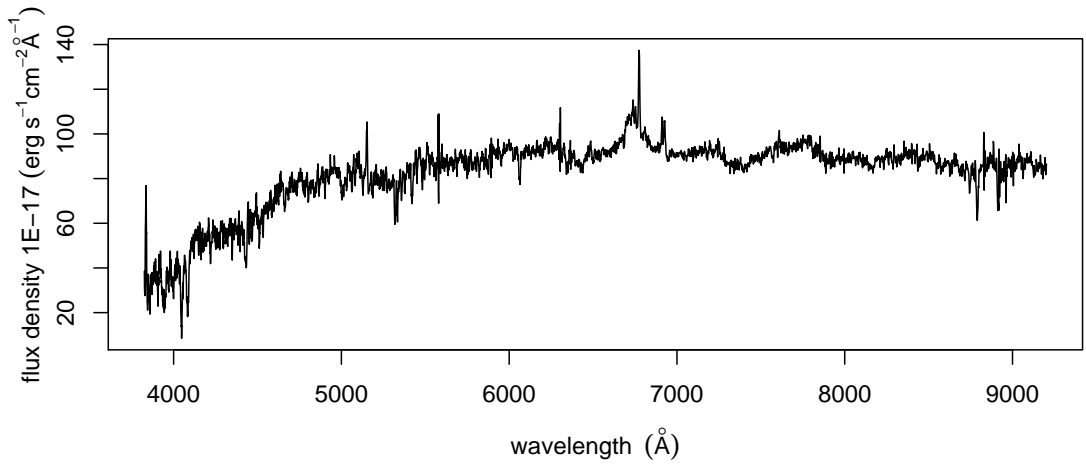}
\caption{SDSS DR9 optical spectrum of NGC 6418.} 
\label{full_spec}
\end{figure*}

\begin{figure}
\plotone{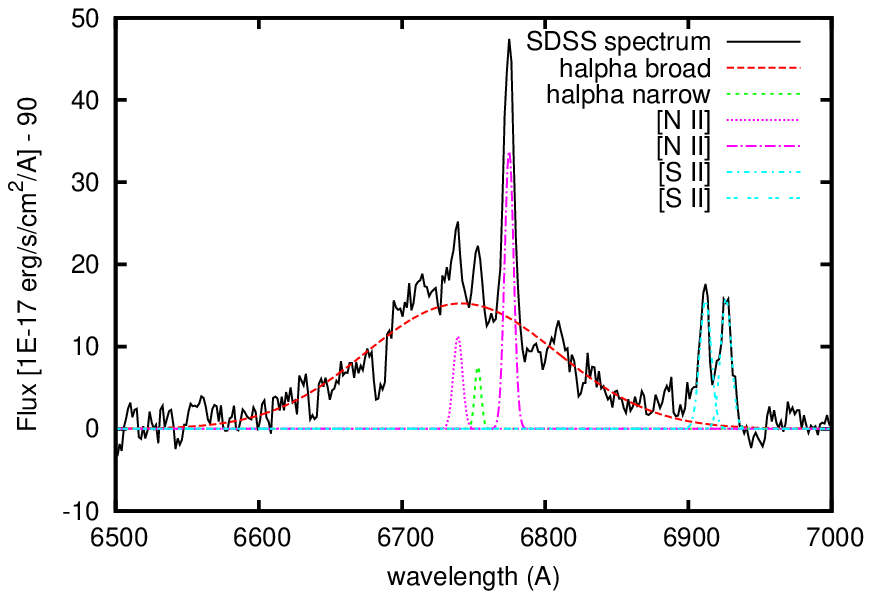}
\caption{ Fit to the $H\alpha$ broad emission line and blended narrow lines in the 
          SDSS DR9 optical spectrum of NGC 6418. The fitted gaussian profiles represent: 
          broad $H\alpha$ (red), [N II]$\lambda\lambda 6548, 83$ (magenta), 
          and [S II]$\lambda\lambda 6717,31$ (cyan) and narrow $H_\alpha$  (green).} 
\label{fig:spec_fit}
\end{figure}

In order to estimate the sublimation radii given by 
equations~\ref{sub_rad} and~\ref{sub_rad_graphite}, it is necessary
to determine $L_{bol}$. However, this is difficult 
to determine accurately for NGC 6418, as the optical spectrum is dominated by 
the stellar continuum and the AGN itself is evidently heavily reddened \citep{Remillard:1993}.

The Sloan Digital Sky Survey (SDSS) optical spectrum 
of NGC 6418 \citep{Ahn:2012} (Figure~\ref{full_spec})
shows broad H$\alpha$ and narrow lines of [OIII]$\lambda 5007$, H$\alpha$, [NII]$\lambda 6548, 6583$ and 
[SII]$\lambda 6717, 6731$, but the 
continuum is dominated by an evolved stellar population. 
The fact that the broad H$\beta$ line is not evident in the spectrum indicates a 
steep broad-line Balmer decrement 
and suggests classification as a Seyfert Type 1.9 
(Sy1.9; \citealt{Osterbrock:1977, Osterbrock:1981}). However, NGC 6418 
is unusual in that the narrow H$\beta$ emission is also very weak (in fact, this line appears 
in absorption) and the [OIII]$\lambda 4959, 5007$ lines are much weaker relative 
to the stellar continuum than is typical in Seyferts, even Sy 1.9s.
Interestingly, these lines are not obviously visible in the earlier (1989) spectrum obtained by
\cite{Remillard:1993}, even though the broad H$\alpha$ line is clearly much 
stronger relative to the narrow H$\alpha$ and [NII] lines than in the SDSS spectrum. 
Evidently, the strong stellar continuum, the foreground reddening and the variable broad 
emission lines make the classification of this source somewhat ambiguous.  \\

To determine the bolometric luminosity of the AGN, we used the relationship established 
between the broad H$\alpha$ luminosity ($L_{bH_\alpha}$) and the bolometric AGN 
luminosity ($L_{AGN}$) in a
large sample of quasars and Sy1 \citep{Richards:2006,Stern:2012}. 

\begin{equation}
L_{bol}=130^{\times2.4}_{\div2.4}\times L_{bH_\alpha}
\label{agn_lum}
\end{equation}

The flux in the broad H$\alpha$ line was measured from 
the SDSS spectrum 
using gaussian profiles to fit and deblend the [N II], [S II] and H$\alpha$ lines. 
In the fit, the wavelengths of the components representing the narrow lines were fixed
at the values determined by the SDSS spectroscopic measurement pipeline ({\em spec1d}; \citealt{Bolton:2012}).
The amplitudes and widths were free parameters, with the exception of [NII]$\lambda 6548$, 
which is constrained so as to preserve its fixed 1:3 intensity ratio with [NII]$\lambda 6583$.
The variances provided by the SDSS spectroscopic data reduction pipeline 
({\em spec2d}; \citealt{Stoughton:2002}) were used to assign weights to each
data point; in addition, we assigned a 10\% systematic error 
to the derived fluxes \citep{Bolton:2012}. The resulting fit is
shown in Figure~\ref{fig:spec_fit} and the parameters derived from the fit are 
summarized in Table~\ref{table1}. The broad H$\alpha$ component has a flux of 
($2563\pm120$) $\times 10^{-17}$
$\text{erg}\;\text{s}^{-1}\text{cm}^{-2}$.
Using this H$\alpha$ flux and assuming a distance of 122 Mpc \citep{Mould:2000},
we calculate the  observed H$\alpha$ broad line luminosity to be
$L^{obs}_{H\alpha} = $($4.56\pm0.85$)$\times10^{40}$ erg$\text{ s}^{-1}$. \\

\begin{figure}
\plotone{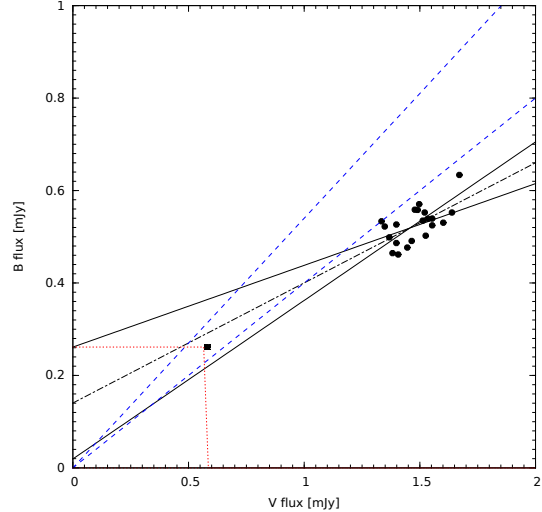}
\caption{ Flux variation gradient diagram of NGC 6418 constructed from observations made at the Fountainwood Observatory in 
Southwestern University. The data are represented by the black dots. The host contribution as indicated by the asterisk 
is 0.58 mJy and 0.29 mJy for the V and B band, respectively. The dashed lines indicate the range of host slopes determined 
in the optical by \cite{Sakata:2010}. The dot-dashed and solid lines indicate the least-square best fit to the range of
the AGN slope.}  
\label{fig:fvg}
\end{figure}

It is clear, however, that a large extinction correction needs to be applied in order to
obtain the intrinsic H$\alpha$ luminosity. From the SDSS spectrum we estimate 
a lower limit to the broad line Balmer decrement of H$\alpha$/H$\beta$ $\ge$ 6.
We used the mean ${H\alpha}/{H\beta}$ from \cite{Dong:2005} and their expression to 
allow for reddening:

\begin{equation}
\log L^{int}_{H\alpha} = \log L^{obs}_{H\alpha} + 1.87(\log({H\alpha}/{H\beta}) - \log(2.97))
\end{equation}

which yields a lower limit to the intrinsic broad H$\alpha$ luminosity of
$L^{int}_{H\alpha}\ge (1.70\pm0.32)\times10^{41}$ erg$\text{ s}^{-1}$. \\

\begin{figure*}
\plotone{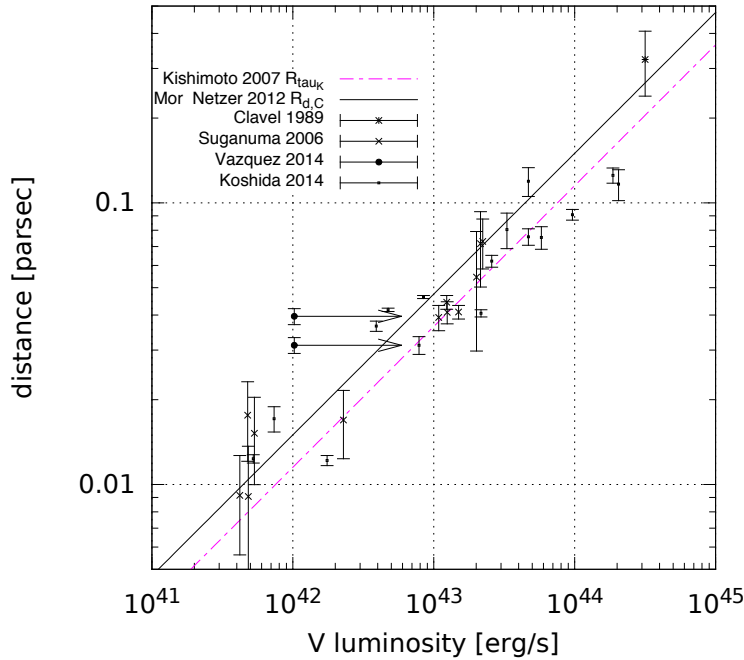}
\caption{ Reverberation lag distance as a function of optical AGN luminosity. The data points are 
          the $K-$band lag measurements of \cite{Koshida:2014,Suganuma:2006,Clavel:1989} and the 3.6 $\mu$m and 4.5 $\mu$m lag measurements of NGC 6418.
          The solid line represents the fit to the ($\tau \propto L^{0.5}$) relationship as found by 
          \citep{Suganuma:2006} and defined as equation 3 in \citep{Kishimoto:2007}.}  
\label{fig:rad_lum}
\end{figure*}

With this lower limit, equation \ref{agn_lum} 
yields a lower limit to the bolometric luminosity of the AGN in
NGC 6418 of $L_{bol}\ge$ ($2.21^{+3.09}_{-1.29}$)$\times10^{43}$ erg$\text{ s}^{-1}$. 
Using \citet{Kaspi:2000}'s relation $ L_{bol} \sim 9 \lambda L_{\lambda}(5100)$ and 
assuming $L_{\lambda}(5500)\sim L_{\lambda}(5100)$ we obtain a lower limit to the 
AGN V-band luminosity of  
$L_{V}\ge$ ($2.46^{+3.43}_{-1.44}$)$\times10^{42}$ erg$\text{ s}^{-1}$. 
For comparison, we used the
flux variation gradient (FVG) method \citep{Choloniewski:1981,Sakata:2010,Haas:2011,Pozo-Nunez:2012,Pozo-Nunez:2014} 
to estimate the (constant) host galaxy contribution within our photometric aperture as 
illustrated in figure \ref{fig:fvg}. Using the B and V fluxes obtained 
from the SU observations (3.5" aperture), we find an $AGN/Host$ ratio of $1.55$, yielding an
an estimate for the AGN contribution to the V-band luminosity of $1.54 \pm 0.53 \times10^{42}\text{ erg s}^{-1}$  
(as reddening corrections have not been applied to the B and V fluxes, this value should be regarded as a lower limit.) 
Thus, within the admittedly large uncertainties, the AGN V-band luminosity estimated from the FVG method is 
consistent with that determined from the H${\alpha}$ luminosity.

Having determined the lower limit on the bolometric luminosity of the AGN, we can determine 
 the dust sublimation radii given by equations~\ref{sub_rad} and~\ref{sub_rad_graphite}.
For silicate dust with a sublimation temperature $\approx 1500$\,K,
 we find $R_{d,Si} \ge$ 
     $60^{+33}_{-21}\times10^{-3}\;\text{pc}$ 
     ($71^{+39}_{-25}$ light days), whereas
for pure graphite dust with sublimation temperature $\sim 1800$\,K, 
 we find  $R_{d,C} \ge$
    $24^{+13}_{-8}\times10^{-3}\;\text{pc}$
    ($28^{+15}_{-10}$ light days). 
    
These sublimation radii bracket the radii derived from the lags at
3.6 $\mu$m ($R_{\tau,3.6} = 31.2^{+2.0}_{-1.9}\times10^{-3}\;\text{pc}$)  
and 4.5 $\mu$m ($R_{\tau,4.5} = 39.5^{+2.6}_{-2.6}\times10^{-3}\;\text{pc}$). 
As $R_{d,Si}$ and $R_{d,C}$ are lower limits, this suggests 
that the bulk of the 3.6 $\mu$m and 4.5 $\mu$m emission comes from the region bounded by the graphite
and silicate sublimation radii, respectively, and is conceivably emitted by the same graphite dust that 
is thought to be responsible for the NIR bump. As already noted, the model graphite dust emission 
spectrum computed by \citet{Mor:2012}, while peaking in
the $2-3 \mu$m range, also emits strongly in the  3.6 -- 4.5 $\mu$m range. Nevertheless, the longer 
lag exhibited by the 4.5 $\mu$m emission implies the presence of a temperature gradient in the emitting region. \\

\renewcommand{\thefootnote}{\alph{footnote}}

\begin{deluxetable}{cccc}
\tablecolumns{6}
\tabletypesize{\scriptsize}
\tablecaption{Emission line fit parameters}
\tablewidth{0pt}
\tablehead{
\colhead{line} & 
\colhead{$\lambda$\tablenotemark{a}} & 
\colhead{\text{flux}\tablenotemark{b}} & 
\colhead{\text{FWHM}\tablenotemark{a}} \\
\colhead{} &
\colhead{(\AA)} &
\colhead{($\text{erg}\;\text{s}^{-1}\text{cm}^{-2}\times 10^{-17}$)} &
\colhead{(\AA)}
}
\startdata
$H_{\alpha\;broad}$ & 6742$\pm$2 & 2563$\pm$120 & 156$\pm$4 \\ 
$H_{\alpha\;narrow}$ & 6753 & 41$\pm$20 & 5$\pm$1 \\ 
$[N II]$ & 6739 & 90$\pm$25 & 8$\pm$1 \\ 
$[N II]$ & 6775 & 269$\pm$26 & 8$\pm$1 \\ 
$[S II]$ & 6912 & 159$\pm$28 & 10$\pm$1 \\ 
$[S II]$ & 6926 & 151$\pm$27 & 9$\pm$1 \\ 
\enddata 

\label{table1}
\end{deluxetable}

\renewcommand{\thefootnote}{\arabic{footnote}}

In K-band reverberation mapping studies of Seyfert 1 galaxies it has been found that the
reverberation radius derived from the time lag is quite tightly correlated with $L_{opt}^{0.5}$, 
where $L_{opt}$ is the AGN
optical luminosity \citep{Suganuma:2006,Koshida:2009,Koshida:2014}. 
This is consistent with the $R\propto L^{0.5}$ relation expected for dust in radiative 
equilibrium. However, \cite{Kishimoto:2007} found that the $K-$band reverberation radii are a 
factor $\sim 3$ smaller than the sublimation radii as predicted by equation~\ref{sub_rad}. 
One possible explanation is that the NIR dust emission is dominated by graphite grains;  
sublimation radii predicted by equation~\ref{sub_rad_graphite} are a factor $\sim 3$ smaller than the Silicate radii 
and thus much closer to the 
$K-$band reverberation measurements (see Fig.~\ref{fig:rad_lum}).
Several other explanations have been advanced for the apparent discrepancy between
the measured dust radii and the sublimation radii predicted for the standard ISM dust composition. 
For example, the dust may include larger grains than the typical size ($a\approx 0.05\mu$m) assumed 
in equation~\ref{sub_rad}\citep{Kishimoto:2007}. \citet{Kawaguchi:2010} investigated the effect
of anisotropic illumination of the torus inner wall by the accretion disk, which permits a smaller torus inner radius
close to the disk plane. Another possibility, proposed by \citet{Pozo-Nunez:2014}, is 
that the torus is very optically thick in the NIR so that only emission from the facing rim of the torus inner wall
is seen, leading to a ``foreshortened'' lag. Modeling of 
the time-dependence of the optical-NIR spectral energy distribution (SED) 
of NGC 4151 by \citet{Schnulle:2013} suggests that the innermost dust is well below the sublimation temperature.  
This implies that the dust is located beyond the sublimation radius, suggesting anisotropic illumination or 
geometrical foreshortening, as envisaged \citet{Pozo-Nunez:2014}.

In Figure~\ref{fig:rad_lum} we plot reverberation radii versus V-band luminosity 
($\lambda L_{\lambda}( \text{V})$) for both the 3.6 $\mu$m and 4.5 $\mu$m lags reported 
here and K-band results taken from \citet{Clavel:1989, Suganuma:2006} and \citet{Koshida:2014}. 
For this purpose, we use the lower limit to the AGN V-band luminosity of NGC 6418 inferred 
from $L^{obs}_{H\alpha}$, as described above.

We also plot \citet{Kishimoto:2007}'s fit 
to the K-band lag data points,

\begin{equation}
R_{\tau,\text{K}} = 0.47\left(\frac{6\lambda L_{\lambda}( \text{V})}{10^{46} \text{ erg\,s}^{-1}}\right)^{1/2} \text{pc}.
\label{RtauK}
\end{equation} 

With the caveat that the NGC 6418 points represent lower limits in luminosity, it can be seen that 
the mid-IR reverberation radii are located above the trend defined by the $K$-band lag times, as expected if 
the 3.6 $\mu$m and 4.5 $\mu$m emission is dominated by cooler dust located somewhat deeper in the torus. 
Equation~\ref{RtauK} predicts $ R_{\tau,\text{K}} \gtrsim 11.6 \times10^{-3}\;\text{pc}$ for NGC 6418, 
given our lower limit on the V luminosity, implying that $R_{\tau,3.6}\lesssim 2.7R_{\tau,\text{K}}$ and 
$R_{\tau,4.5}\lesssim 3.4R_{\tau,\text{K}}$, respectively.

For dust grains in radiative equilibrium, the radius at which grains have a temperature $T$ is approximately, 

\begin{equation}
\frac{R_d}{R_{sub}} \simeq \left(\frac{T}{T_{sub}}\right)^{\alpha} \label{grain_rad},
\end{equation}

where $R_{sub}$ is the sublimation radius and $\alpha \approx 2 - 2.8$ depends on the dust composition. 
In combination with Wien's Law, Equation~\ref{grain_rad} provides a
rough estimate of the largest radius at which the dust contributes to the torus emission at a 
specific wavelength. For the typical ISM composition of Equation~\ref{sub_rad} ($\alpha = 2.6$),
we find $R_{3.6}/R_{\text{K}} \simeq 3.6$, $R_{4.5}/R_{\text{K}} \simeq 6.4$ and 
$R_{4.5}/R_{3.6} \simeq 1.8$. The values for $R_{3.6}/R_{\text{K}}$ and $R_{3.6}/R_{\text{K}}$ 
exceed the empirical upper limits determined from reverberation mapping, while the value 
of $R_{4.5}/R_{3.6}$ agrees with the ratio of the reverberation lags 
($R_{\tau,4.5}/R_{\tau,3.6} = 1.3\pm0.7$). 

However, in clumpy torus models\citep[e.g.][]{Nenkova:2008a,Nenkova:2008b}, there is 
a wide range of dust temperature within a typical cloud, which therefore emits 
a broad IR spectrum. 
In the models of \cite{Nenkova:2008b}, the bulk of the emission at $\lambda \lesssim  5 \mu$m 
emerges from clouds at no more than twice the inner radius (see \cite{Nenkova:2008b} their Fig. 13). 
Thus, the relative sizes of the reverberation radii at 3.6 $\mu$m, 4.5 $\mu$m and $K-$band seem
consistent with at least some clumpy tori models.

It is well established, mainly from Balmer line reverberation mapping 
(\citealt{Bentz:2013, Greene:2010} and references therein) that the 
broad emission line region follows a similar $R\sim L^{1/2}$ size-luminosity 
relationship. For a given AGN luminosity, the BLR reverberation radius is a 
factor $4-5$ smaller than the K-band dust emission reverberation radius \citep{Suganuma:2006, Koshida:2014}, 
as expected in the AGN unification paradigm. Interestingly, radii derived 
from \cite{Mor:2012}'s SED fits suggest that the NIR emission component 
attributed to hot graphite dust clouds occupies a region intermediate between the 
BLR and K-band reverberation radii \citep[see][Figure~13]{Koshida:2014}, 
consistent with the idea that this dust resides in the outer BLR clouds. In their 
analysis of mid-IR (12$\mu$m) interferometric observations, \citet{Burtscher:2013} find
that although source sizes scale in a similar way with luminosity, there is a much 
larger scatter, with mid-IR source radii ranging from $\lesssim 4$ to $20\times R_{\tau,\text{K}}$. 
A clearer picture of the structure of the AGN emission regions beyond the accretion 
disk is therefore beginning to emerge. Placing our results in this context, the 
reverberation radii derived from the 3.6 $\mu$m and 4.5 $\mu$m light curves are 
consistent with the variable emission at these wavelengths arising in the inner clouds 
of the torus. However, we note as a caveat that NGC 6418 exhibits an atypical optical 
spectrum for a Seyfert 1, with a relatively strong, broad H$\alpha$ line but with a steep 
Balmer decrement, relatively weak narrow lines (for instance, the equivalent width of 
[OIII]$\lambda 5007$ is only $\sim 3$\AA, that of narrow H$\alpha \sim 0.5$\AA) and with 
stellar emission dominating the optical continuum. This indicates that the BLR and AGN 
UV-optical continuum are subject to heavy extinction along the line-of-sight, raising the 
possibility that the circum-nuclear dust distribution may be more quasi-spherical than 
toroidal in nature. 

\section{Summary}
\label{sec:Summary}

We have presented initial results from the first year of a two-year campaign of IR (3.6 $\mu$m and
 4.6 $\mu$m) and optical (B and V) monitoring  
of a sample of 12 Seyfert 1 galaxies using the Spitzer Space Telescope supported by ground-based optical 
observations. In NGC 6418,   
we have found a lag between the 
mid-IR and optical light curves, 
with a time delay of $37.2^{+2.4}_{-2.2}$ days ($31.2^{+2.0}_{-1.9}\times10^{-3}\;\text{pc}$)
at 3.6 $\mu$m 
and $47.1^{+3.1}_{-3.1}$ days ($39.5^{+2.6}_{-2.6}\times10^{-3}\;\text{pc}$) 
at 4.5 $\mu$m,
respectively. The 3.6 $\mu$m emission
leads the 4.5 $\mu$m emission by  
$13.9^{+0.5}_{-0.1}$ days ($11.7^{+0.4}_{-0.1}\times10^{-3}\;\text{pc}$). 
These results indicate that the dust emitting the bulk of 
the 3.6 $\mu$m and  4.5 $\mu$m emission 
is located at a distance $\approx 1$ light-month ($\approx 0.03$\,pc) from
the source of the AGN UV--optical continuum.  \\

The nucleus of NGC6814 appears to be heavily reddened, with a
broad line Balmer decrement of H$\alpha$/H$\beta$ $\ge$ 6. For this reason,
we can only determine a lower limit for the intrinsic luminosity of the AGN and hence 
lower limits on the dust sublimation radii. The reverberation radii are a
factor $\sim 2$ smaller than the sublimation radius lower limit for silicate grains
(sublimation temperature  $\approx 1500$ K; $R_{d,Si} \ge 60^{+33}_{-21}\times10^{-3}\;\text{pc}$),
but consistent with that for pure-graphite 
grains (sublimation temperature $\approx 1800$ K;  $R_{d,C} \ge 24^{+13}_{-8}\times10^{-3}\;\text{pc}$). 
Reverberation radii derived from K-band variability studies of other Seyferts are similarly 
a factor $\sim 3$ smaller than the silicate sublimation radius. 
It seems possible that some of the emission in the 3.6 -- 4.5 $\mu$m range comes from
hot graphite dust located within the region bounded by $R_{d,C}$ and $R_{d,Si}$, 
whose presence is suggested by SED model-fitting.

The 3.6 and 4.5 $\mu$m reverberation radii fall above the extrapolated K-band size-luminosity 
relationship by factors $\lesssim 2.7$ and $\lesssim 3.4$, respectively, while the 4.5 $\mu$m 
reverberation radius is only 27\% larger than the 3.6 $\mu$m radius. This indicates a steeper
temperature gradient than expected for optically thin dust in radiative equilibrium but is 
consistent with clumpy torus models, in which individual optically thick clouds emit strongly 
over a broad wavelength range.

\acknowledgments

We dedicate this paper to the memory of our great friend, colleague and mentor David Axon, 
who initiated  
this project and brought  the collaboration together.
We would like to thank Davide Lena for improvements to the manuscript and  
discussions on spectroscopic data reduction. SU thanks Lex Shaw for a generous instrumentation donation.
This work is based [in part] on observations made with the Spitzer Space Telescope, 
which is operated by the Jet Propulsion Laboratory, California Institute of Technology under 
a contract with NASA. Support for this work was provided by NASA through an award issued 
by JPL/Caltech, GO-80120. 
This research has made use of the SIMBAD database,
operated at CDS, Strasbourg, France,
and of the NASA/IPAC Extragalactic Database (NED) which is operated by the Jet Propulsion Laboratory, 
California Institute of Technology, under contract with the National Aeronautics and Space Administration.

Funding for SDSS-III has been provided by the Alfred P. Sloan Foundation, the Participating 
Institutions, the National Science Foundation, and the U.S. Department of Energy Office of 
Science. The SDSS-III web site is http://www.sdss3.org/. 

SDSS-III is managed by the Astrophysical Research Consortium for the Participating Institutions of 
the SDSS-III Collaboration including the University of Arizona, the Brazilian Participation Group, 
Brookhaven National Laboratory, Carnegie Mellon University, University of Florida, the French 
Participation Group, the German Participation Group, Harvard University, the Instituto de 
Astrofisica de Canarias, the Michigan State/Notre Dame/JINA Participation Group, Johns Hopkins 
University, Lawrence Berkeley National Laboratory, Max Planck Institute for Astrophysics, 
Max Planck Institute for Extraterrestrial Physics, New Mexico State University, New York University, 
Ohio State University, Pennsylvania State University, University of Portsmouth, 
Princeton University, the Spanish Participation Group, University of Tokyo, University of Utah, 
Vanderbilt University, University of Virginia, University of Washington, and Yale University. 

The Liverpool Telescope is operated on the island of La Palma by Liverpool John Moores 
University in the Spanish Observatorio del Roque de los Muchachos of the Instituto de 
Astrofisica de Canarias with financial support from the UK Science and Technology 
Facilities Council. 

The Faulkes Telescope Project is an educational and research arm of the Las Cumbres
Observatory Global Telescope Network (LCOGTN).

BMP is supported by NSF grant AST-1008882.

\bibliographystyle{apj}
\bibliography{ngc6418_reverb}

\newpage

\appendix

\section{Appendix A - Ensemble Photometry}
\label{app:a}

We begin by defining a region within each image containing NGC 6418 
and several nearby reference stars. 
Next, we determine a background value for this region by fitting a 
gaussian to the histogram of pixel 
values: the peak yields the background 
value and the width its uncertainty. 
Sources are detected using the STARS program of XVISTA, which employs an 
algorithm based on the FIND procedure within DAOPHOT \citep{Stetson:1987}. 
Candidate objects which survive cuts 
in several parameters such as full-width at 
half-maximum, sharpness and roundness, are selected for 
aperture photometry. 
We measure the brightness of each object using the PHOT program in XVISTA, 
which sums all counts within a circular aperture, 
including weighted contributions from pixels that  
lie partially inside the aperture. 
PHOT also measures the 
median pixel value within an annulus around each object 
to determine a local sky value
and subtracts this from the object counts. 
Finally, the remaining object counts are converted to an instrumental magnitude. \\

The second stage of the analysis subjects the 
measured instrumental magnitudes to inhomogeneous ensemble 
photometry \citep{Honeycutt:1992}. Small differences in sky brightness, transparency,
exposure time, and other factors can cause all objects in some particular 
exposure to appear 
slightly brighter or dimmer than average; 
ensemble photometry is designed to identify these
systematic changes and remove their effects.

\begin{figure}
\plotone{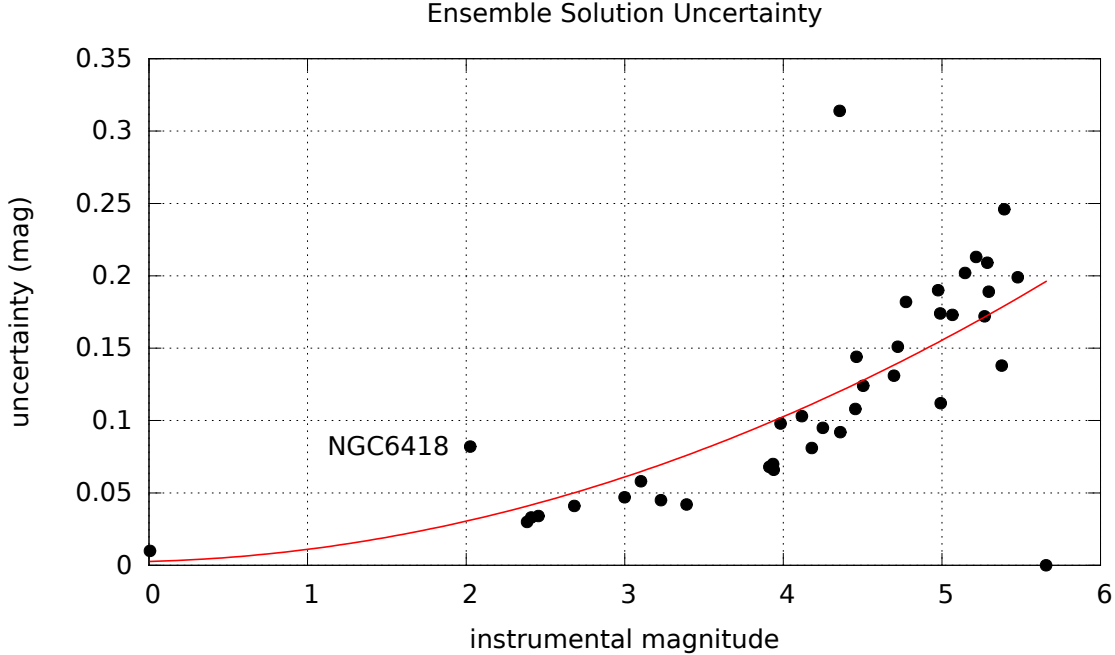}
\caption{Spitzer ch1 (3.6 $\mu$m) uncertainties for the ensemble solution. The curve in 
red is the least-square fit to a quadratic function.}
\label{fig:unc}
\end{figure}

 \cite{Honeycutt:1992} defines the equation of condition as
\begin{equation}
m(e,s)=m0(s) + em(e) \text{,}
\end{equation}
where ${m(e,s)}$ is the instrumental magnitude of star ${s}$ in exposure ${e}$
and ${m0}$ is the intrinsic instrumental magnitude of that star.
The ``exposure magnitude'', ${em}$, of an image 
accounts for variations in extinction, 
exposure time, background intensity and 
other effects that are common to all sources in an image. 
We note that even without the transparency issues 
that are typical of ground observations, the Spitzer IR data will have small 
variations due to changes in orientation and 
background illumination of the space telescope. The quantity that we want to minimize is
\begin{equation}
\beta=\sum\limits_{e=1}^{ee}\sum\limits_{s=1}^{ss}[m(e,s)-m0(s)-em(e)]^{2}w(e,s) \text{,}
\end{equation}
where ${w(e,s)}$ is the weight of each instrumental magnitude; 
we take its value to be
$\sigma(m(e,s))^{-2}$. 
This technique yields the best fit value of $m0(s)$ for each source, 
assuming no intrinsic variability, 
and an empirical estimate of the uncertainty. 
In an ideal experiment, the uncertainty would be equal to that derived from
the quadrature sum of the shot noise of the source, 
the sky noise and the detector read noise. 
This empirical estimate of the uncertainty is valid only for constant sources,
such as the reference stars, 
but not for sources which vary intrinsically from one  image to the next. 
The estimated uncertainties for the Spitzer 
Channel 1 data are shown in Figure \ref{fig:unc}
as a function of instrumental magnitude, with a quadratic fit to the  
reference stars in the field.

\section{Appendix B - Cross-Correlation Codes Results}
\label{app:b}
In an effort to give a comprehensive picture of the results obtained by different software packages
used to determine the lag time between light curves we have included this appendix with table \ref{dtable} of all 
results. The table contains the analyses of individual and combined optical datasets versus the infrared channels
of the Spitzer Space Telescope. The first column indicates the Spitzer channel. The table has three sections, one for each
 of the software packages we used. Columns 2 through 6 are values obtained for our in-house cross-correlation package.
 Of those, columns 2-4 represent the difference from the median to the 25\% value of the interquantile range (IQR), 
 the median of the distribution and the difference from the median to the 75\% value of the IQR, respectively. 
 Columns 5 and 6 are the mean and the standard deviation. Column 7 is 
 the mean and the standard deviation for Peterson's code \citep{Peterson:2004}. Columns 8-10 are Zu's \citep{Zu:2011} corresponding 
 to the low, mid and high
  values of the lag.
  
\begin{deluxetable}{cccccccccc}
\tabletypesize{\scriptsize}
\tablecaption{Comparison of cross-correlation methods}
\tablewidth{0pt}
\tablehead{
\colhead{} & &  \colhead{Vazquez}  & & & \colhead{Peterson} & & \colhead{Zu} \\
\colhead{channel} & 
\colhead{IQR 25\%} & 
\colhead{median} & 
\colhead{IQR 75\%} &
\colhead{mean $\pm$ std} &  
\colhead{mean $\pm$ std} & 
\colhead{low} &
\colhead{mid} &
\colhead{high} &
\colhead{dataset}
}
\startdata
ch1 & -5.1 & 42.7 & 5.3 & 42.6 $\pm$ 7.8 & 42.3 $\pm$ 9.6 & -0.7 & 40.9 & 1.2 & FTN ISIS \\[4pt]
ch2 & -3.1 & 50.4 & 3.2 & 50.5 $\pm$ 7.1 & 52.2 $\pm$ 13.5 & -6.5 & 53.2 & 5.2 & FTN ISIS \\[4pt]
ch1 & -2.4 & 35.0 & 2.2 & 35.1 $\pm$ 3.7 & 35.1 $\pm$ 4.2 & -0.8 & 33.5 & 6.9 & LT ISIS \\[4pt]
ch2 & -2.7 & 47.3 & 3.5 & 47.8 $\pm$ 4.0 & 48.9 $\pm$ 4.4 & -1.2 & 50.1 & 0.6 & LT ISIS \\[4pt]
ch1 & -5.5 & 42.6 & 5.4 & 42.2 $\pm$ 8.1 & 47.5 $\pm$ 10.2 & -1.3 & 39.6 & 21.1 & FTN XVISTA \\[4pt]
ch2 & -3.6 & 50.6 & 3.0 & 50.2 $\pm$ 7.1 & 60.2 $\pm$ 19.2 & -31.1 & 69.0 & 1.2 & FTN XVISTA \\[4pt]
ch1 & -2.4 & 27.1 & 2.8 & 27.5 $\pm$ 4.1 & 28.0 $\pm$ 6.0 & -0.7 & 29.2 & 0.8 & LT XVISTA \\[4pt]
ch2 & -2.6 & 36.1 & 3.3 & 36.5 $\pm$ 3.7 & 35.2 $\pm$ 4.2 & -2.7 & 33.2 & 0.7 & LT XVISTA \\[4pt]
ch1 & -2.4 & 34.5 & 2.1 & 34.4 $\pm$ 3.6 & 34.5 $\pm$ 3.9 & -8.2 & 35.6 & 1.0 & LT + FTN XVISTA \\[4pt]
ch2 & -3.7 & 44.6 & 3.5 & 44.5 $\pm$ 4.2 & 42.2 $\pm$ 4.6 & -2.7 & 37.8 & 6.8 & LT + FTN XVISTA \\[4pt]
ch1 & -2.2 & 37.2 & 2.4 & 37.2 $\pm$ 3.3 & 36.7 $\pm$ 3.4 & -6.5 & 40.4 & 0.7 & LT + FTN ISIS \\[4pt]
ch2 & -3.1 & 47.1 & 3.1 & 47.3 $\pm$ 4.6 & 48.6 $\pm$ 3.7 & -4.7 & 49.5 & 1.2 & LT + FTN ISIS \\[4pt]
ch1/ch2 & -0.1 & 13.9 & 0.5 & 14.0 $\pm$ 0.7 & 14.6 $\pm$ 6.0 & -2.9 & 13.2 & 5.8 & CH1 vs CH2  \\[4pt]
\enddata 

\label{dtable}
\end{deluxetable}  

\begin{figure}
\centering
\subfigure[CH1 vs CH2]{%
  \includegraphics[width=.4\textwidth]{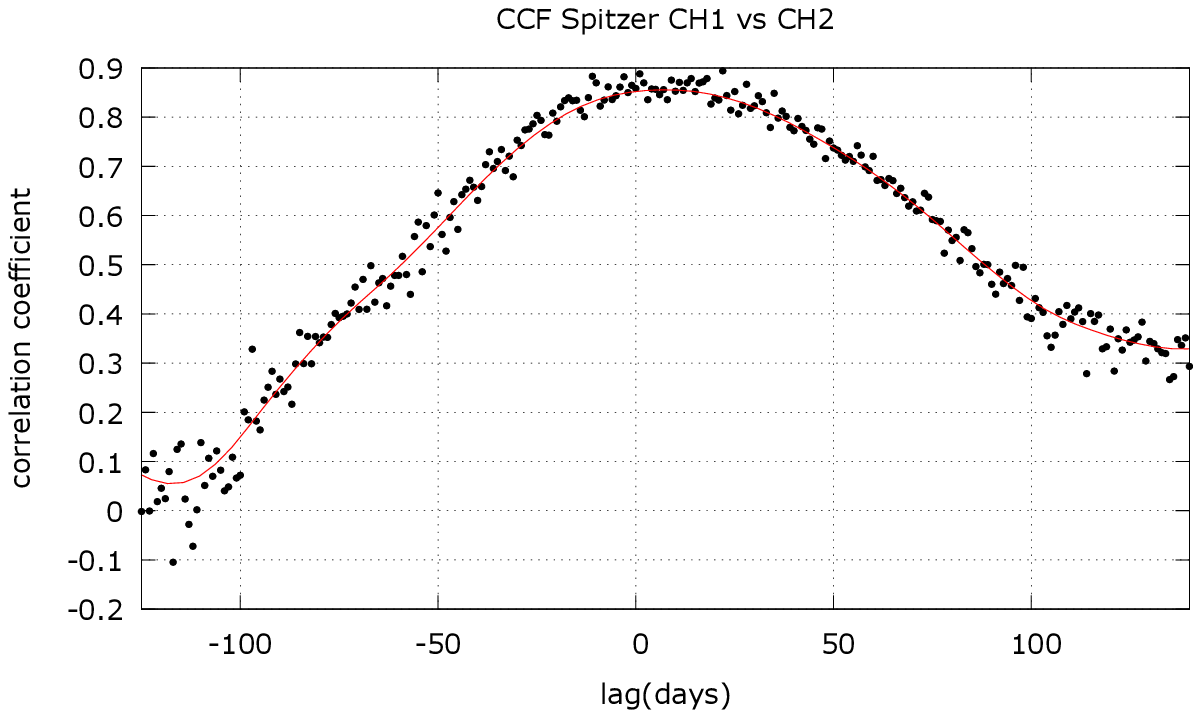}
  }
\subfigure[Optical vs CH1]{%
  \includegraphics[width=.4\textwidth]{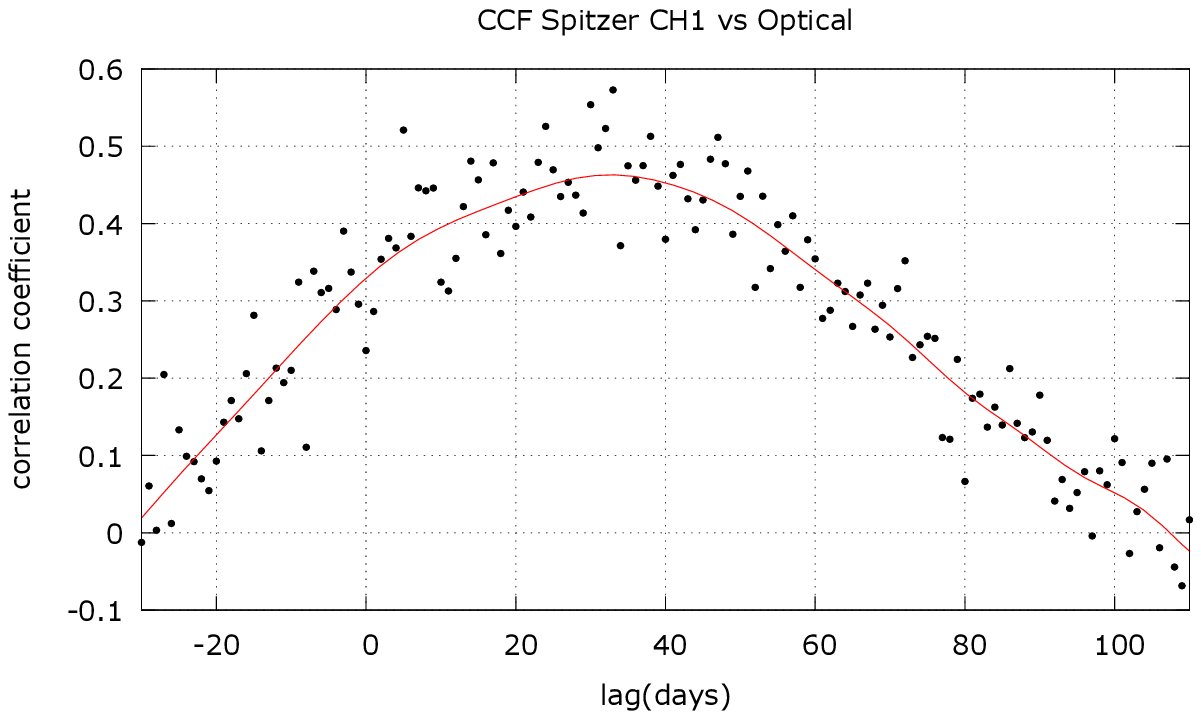}
  }
\subfigure[Optical vs CH2]{%
  \includegraphics[width=.4\textwidth]{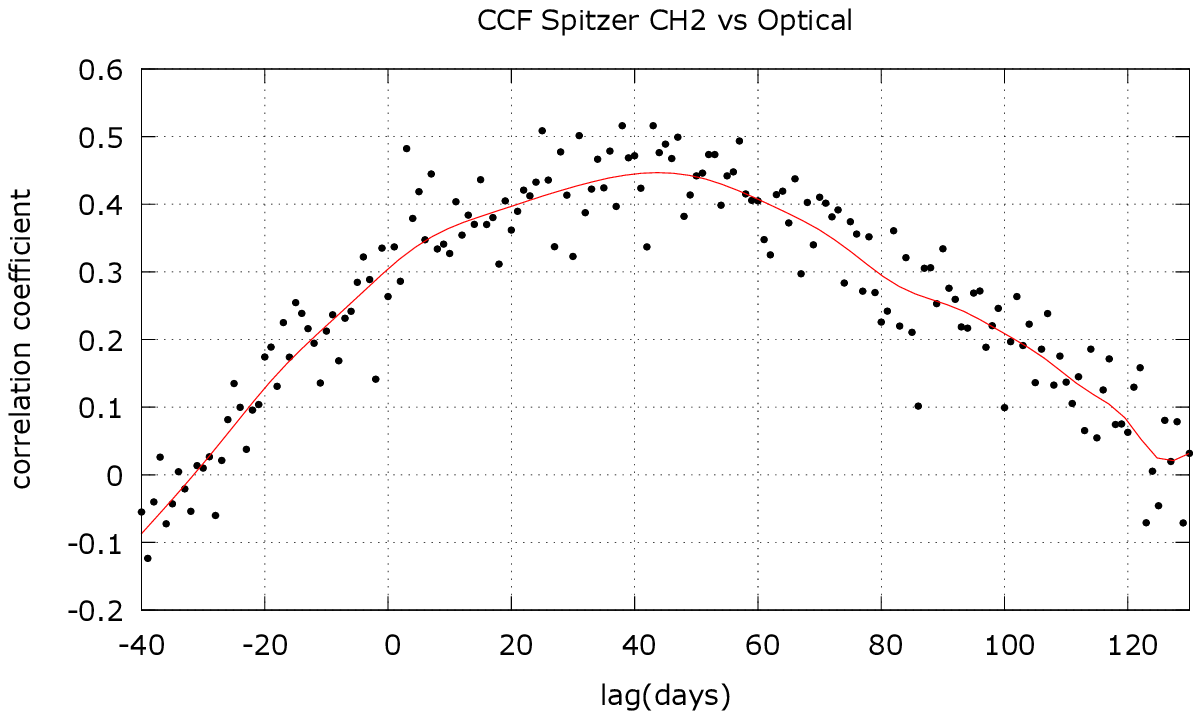}
 }
\caption{Sample realizations of the CCFs for each pair of light curves. The fitted line is a cubic spline.}
\label{fig:ccf}
\end{figure}
 
\section{Appendix C - Cross-Correlation Function and the Cross-Correlation Centroid Distribution }
\label{app:c}

Our simulations employ one thoudsand realizations of the light curves. 
Each synthetic light curve 
is generated by replacing each magnitude measurement with 
an artificial datum. This consists of the measured magnitude plus a random deviate 
drawn from a gaussian distribution with a mean of zero and standard deviation equal to 
the uncertainty in the measured value. We compute the CCFs and the corresponding weighted 
mean lags for each set of synthetic light curves to construct a distribution 
of the CCF centroids, the CCCD. \\

The CCFs are often not symmetrical functions,
and the skewness of these functions affects the calculation
of their centroids.
The question is -- how to select the significant portion of each distribution,
while discarding the uninteresting wings?
Figure \ref{fig:ccf} shows representative single realizations of the CCFs;
it is obvious that the centroid of each CCF will depend on the range of
data chosen for further calculation.
In this work, we have 
adopted an algorithm that uses properties of each distribution itself
to select the subset of measurements for the centroid calculation.
First, we fit a cubic spline to the distribution
in each realization,
and compute the standard deviation, $\sigma$,  
between the spline and the data.
We adopt 2$\sigma$ as a measure of the dispersion within the CCF.
We set a threshold in correlation which is 
the peak of the CCF minus this dispersion:
$K = {\rm peak} - 2\sigma$.
All the CCF values greater than $K$ 
are then used to calculate the centroid of that particular CCF.
The fitted spline is shown together with the computed CCF($\tau$) data points. 
We found that 
for optical vs 3.6 $\mu$m, the top 24\% of CCF data was used, for the optical vs. 4.5 $\mu$m 
the top 23\%, and for the 3.6 $\mu$m vs 4.5 $\mu$m the top 6\%. The threshold clearly is dependent on the 
noise characteristics of the underlying light curves which explains why the Spitzer light curves
have a smaller data percentage used in the centroid calculation. \\

After calculating the centroid of each realization of the CCF in this 
manner, we then combine all the centroids 
to create the cross-correlation centroid distribution (CCCD)
for that pair of light curves.
We choose the median value in the CCCD 
as the time lag between the two light curves. \\

\end{document}